\documentclass[twocolumn,aps,prl,superscriptaddress]{revtex4-1}
\usepackage[latin9]{inputenc}
\usepackage{amsmath}
\usepackage{amssymb}
\usepackage{graphicx}
\usepackage{ulem}
\usepackage[colorlinks,linkcolor=blue,anchorcolor=blue,citecolor=blue,urlcolor=blue]{hyperref}
\makeatletter

\usepackage{color}
\usepackage{bm}
\usepackage{float}

\newcommand{\vb}{\textbf{\textit{v}}}
\newcommand{\Eb}{\textbf{\textit{E}}}
\newcommand{\Bb}{\textbf{\textit{B}}}

\makeatother

\begin{document}
\title{
Relativistic Topological Waves from Cherenkov and Doppler Resonances \\
in Self-Magnetized Laser Plasmas
}

\author{Xiaofei Shen}

\affiliation{Institut f\"ur Theoretische Physik I, Heinrich-Heine-Universit\"at D\"usseldorf,
	40225 D\"usseldorf, Germany}

\author{Lars Reichwein}

\affiliation{Institut f\"ur Theoretische Physik I, Heinrich-Heine-Universit\"at D\"usseldorf,
	40225 D\"usseldorf, Germany}

\author{Alexander Pukhov}
\email{pukhov@tp1.uni-duesseldorf.de}


\affiliation{Institut f\"ur Theoretische Physik I, Heinrich-Heine-Universit\"at D\"usseldorf,
	40225 D\"usseldorf, Germany}

\date{\today}

\begin{abstract}
 
Strong magnetic fields at plasma-plasma interfaces can be naturally produced in laser-plasma interactions. 
Using theoretical analysis and fully three-dimensional particle-in-cell simulations, we demonstrate that relativistic  topological waves can be generated via Cherenkov and Doppler resonances  
in the interaction of intense femtosecond laser pulses with near-critical-density plasmas.  
At the self-magnetized plasma-plasma interface, a new slow-wave branch appears. Its phase velocity is much smaller than the group velocity of the laser pulse and the electron beam velocity. Therefore, the Cherenkov resonance condition can be easily satisfied. Furthermore, since electrons undergo betatron oscillations, Doppler resonances may also occur and are responsible for the excitation of several frequency-shifted branches observed in our simulations. 
After the passage of the laser pulse, we observe a fast remnant mode with relativistic amplitude and frequency close to the local plasma frequency. This mode continues to accelerate electrons further for many tens of laser periods even after the laser pulse has left the plasma.   

\end{abstract}

\maketitle

Strong magnetic fields generated in laser-plasma interactions (LPIs) have been observed for many decades \cite{Stamper1971,Raven1978}.  
The initial interest was motivated by their significant impact on inertial confinement fusion (ICF), such as surface energy transport and anomalously fast plasma blowoff \cite{Jones1983,Yu1985,Yu1989}. 
With the rapid development of laser technology, 
magnetic fields with strength about 10 MegaGauss (MG) have been produced in experiments by irradiating capacitor-coils with nanosecond pulses \cite{Santos2015,Law2016}. As an external quasistatic source, such strong magnetic fields may improve ICF yields \cite{Slutz2012}, provide new capabilities for high-energy-density physics and laboratory astrophysics \cite{Yao2021}, and manipulate pico- or femtosecond LPI processes \cite{Wang2015,Arefiev2016,Li2018}.  
On the other hand, when a relativistic laser pulse with intensity far exceeding $10^{18}\,{\rm W/cm^2}$ interacts with plasmas, the amplitude of self-generated magnetic fields can reach tens of MG to GigaGauss (GG) scale which is strong enough to magnetize energetic electrons in picoseconds or even shorter timescales \cite{Borghesi1998,Pukhov1999,Sarri2012,Liu2013,Nakatsutsumi2018}.  
Extensive studies have been performed to investigate the laser propagation \cite{Weng2017,Shi2018}, particle acceleration \cite{Liu2013,Arefiev2016,Nakatsutsumi2018} and generation of secondary sources \cite{Wang2015,Shen2021,Kaymak2016} in magnetized plasmas.

However, most of these studies have overlooked two potentially important factors: the resonant excitation of a rich variety of waves in magnetized plasmas,  
and its feedback effect on particle acceleration and radiation.  
Only some works, based on an assumption of uniform plasmas, have discussed THz radiation from laser wakefield acceleration with an external DC magnetic field applied perpendicular to the laser propagation direction. There a resonance occurs at the intersection of the Cherenkov condition and lower branch of extraordinary mode \cite{Yoshii1997,Hu2013,Tailliez2022}.  
Recently, cold magnetized plasmas have been investigated as topological materials. At a plasma-vacuum or plasma-plasma interface, various topological modes (surface plasma waves, SPWs) have been demonstrated to exist, where the direction of wave propagation is parallel to that of the constant background magnetic field or has a finite crossing angle \cite{Silveirinha2015,Parker2020,Fu2021,Rajawat2022}. An external electromagnetic source is applied to excite the edge modes \cite{Fu2021,Rajawat2022}. 

In intense LPIs, in principle, such or even more abundant topological waves may be generated without requiring external magnetic fields or wave sources, since the self-generated magnetic fields are sufficiently strong and have diverse topologies \cite{Liu2013,Nakatsutsumi2018,Shen2021,Jiang2021},  and plasma-vacuum or plasma-plasma interfaces are naturally produced \cite{Liu2013,Nakatsutsumi2018}. More importantly, driven by intense laser pulses or high-flux energetic electron beams, it might be possible to excite  topological waves with relativistic amplitudes that further have strong influences on the wave propagation in plasmas and energy coupling to particles and radiation. 
However, until now, little attention has been paid to them, 
likely because of too rich modes including strong laser and quasistatic modes existing simultaneously in intense LPIs.

In this Letter, we, for the first time, demonstrate  
that relativistic magnetized topological plasma waves (mTPWs) can be naturally excited 
in intense LPIs and investigate their effect on further electron acceleration, where a tightly-focused intense femtosecond laser pulse interacts with a near-critical-density (NCD) plasma. Our theoretical analysis reveals that at the magnetized plasma-plasma interface, a new slow-wave branch can be excited. Since its phase velocity is much smaller than the laser group velocity and the electron beam velocity, strong Cherenkov radiation occurs. Furthermore, due to the betatron motion of electrons, it is possible to excite normal and anomalous Doppler resonances \cite{Martin2002,Kartashova2009}. Our three-dimensional (3D) particle-in-cell (PIC) simulations clearly verify the existence of these novel modes at the interface of a low-density channel plasma and high-density filament plasma, where a strong self-generated magnetic field with amplitude up to GG scale and direction perpendicular to the laser propagation direction is generated. We observe a long-lasting fast mode with relativistic amplitude and frequency close to local plasma frequency. Electrons are further accelerated by this strong electromagnetic wave after the laser pulse has left.

\begin{figure}
	\includegraphics[width=6.5cm]{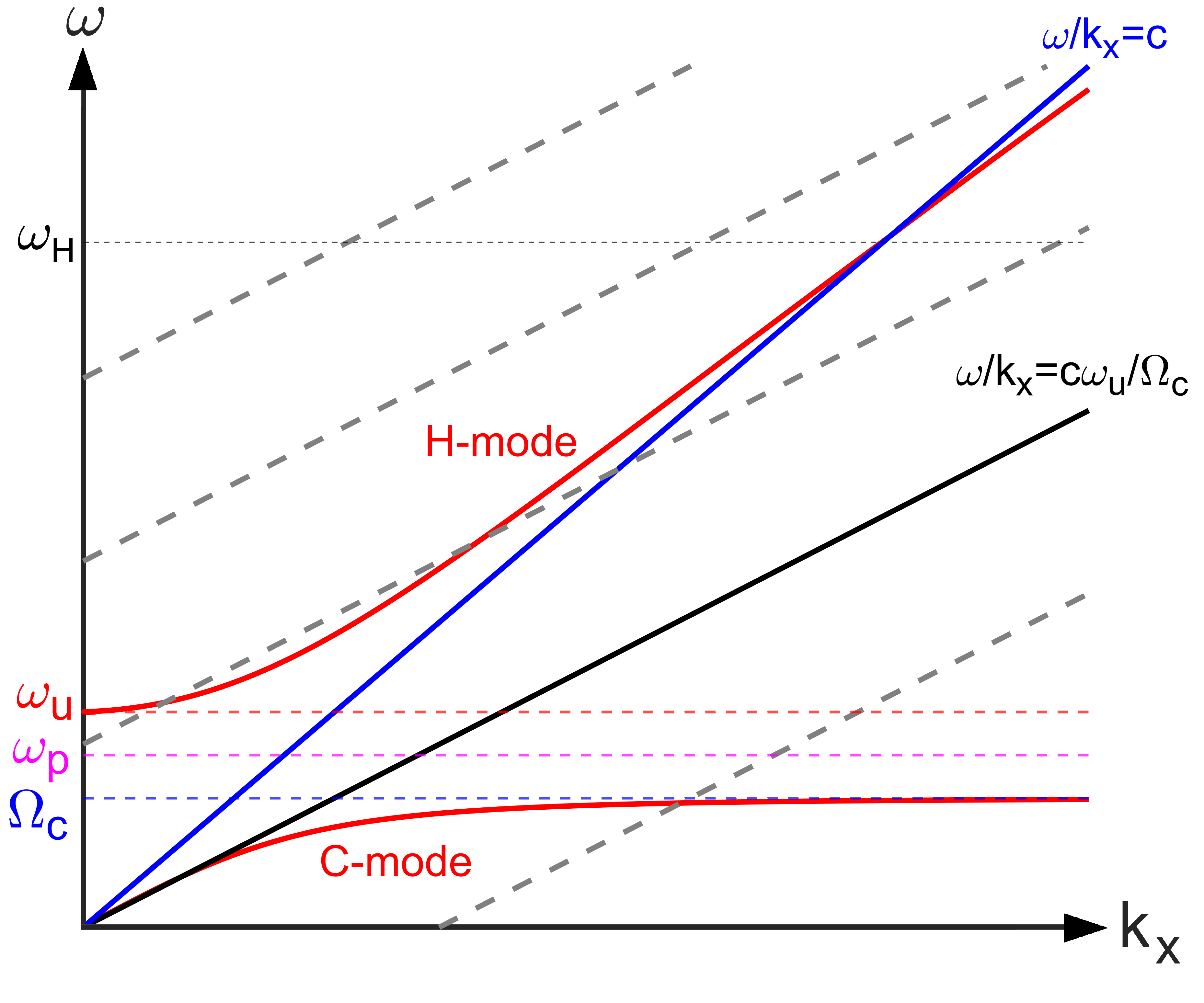}
	\caption{Dispersion curves of the magnetized waves. The red solid lines illustrate the dispersion relation determined by  Eq. (\ref{dispersion}), while  
	the solid blue and black lines represent the Cherenkov condition with different longitudinal velocities. The gray dashed lines visualize the Doppler resonances.
		\label{fig:fig1}
	}
\end{figure}

To derive the dispersion relation of topological waves at a magnetized plasma-plasma interface, we adopt a cold plasma model and consider a geometry where a constant background magnetic field $B_0$ exists in  $z$-direction and the plasma interface at  $y=0$ plane separates two semi-infinite plasmas with electron density $n_{e1}$ and $n_{e2}$. The linearized fluid equations are \cite{Uberoi1975,Stix1992}

\begin{eqnarray}
	\partial_t{\vb} &=& e{\Eb}/m_e + \Omega_c{\vb\times \hat{\textbf{\textit{z}}}},\\
	\partial_t{\Eb}	 &=&  c\nabla\times{\Bb} - m_e\omega_p^2{\vb}/e,\\
	\partial_t{\Bb} &=& -c\nabla\times{\Eb},
\end{eqnarray}
where $\vb$, $\Eb$, $\Bb$ are perturbed velocity, electric and magnetic field, $\omega_p = \sqrt{4\pi n_ee^2/m_e}$ is the plasma frequency, $\Omega_c=eB_0/m_ec$ is the cyclotron frequency, $e$ and $m_e$ are electron charge and mass, and $c$ is light speed.

After some algebra and considering the boundary conditions, we have \cite{Uberoi1975}
\begin{eqnarray}
	k_x^2 + k_{y1,y2}^2 - \frac{\omega^2}{c^2}\varepsilon_{1,2} = 0,\hspace{65pt}&\label{wave}\\
	\varepsilon_1\sqrt{k_x^2 - \frac{\omega^2}{c^2}\varepsilon_2} + \varepsilon_2\sqrt{k_x^2 - \frac{\omega^2}{c^2}\varepsilon_1} = k_x\left(\varepsilon_1\eta_2 - \varepsilon_2\eta_1\right),\hspace{-1pt}&\label{dispersion}
\end{eqnarray}
where 
\begin{eqnarray}
\varepsilon_{1,2} &=& 1 - \frac{\omega_{p1,p2}^2}{\omega^2}\frac{\omega^2 - \omega_{p1,p2}^2}{\omega^2 - \omega_{p1,p2}^2 - \Omega_c^2},\\ 
	\eta_{1,2} &=& -\frac{\omega_{p1,p2}^2\Omega_c}{\omega(\omega^2 - \omega_{p1,p2}^2 - \Omega_c^2)}. 
\end{eqnarray}
Eqs. (\ref{wave}) and (\ref{dispersion}) do not have simple solutions. However, under certain approximations, we can have some useful insights. (i) When $\omega\rightarrow0$, $\varepsilon,\eta\rightarrow\infty$, and by considering a case with $n_{e1}\gg n_{e2}$, we have $ck_x/\omega\approx\omega_{u}/\Omega_c$, where $\omega_{u}=\sqrt{\omega_{p2}^2 + \Omega_c^2}$ is the upper hybrid frequency \cite{Stix1992}. Therefore, its phase velocity $v_{ph}=\omega/k_x \approx (\Omega_c/\omega_{u})c$ can be much smaller than the light speed for strongly magnetized plasmas. The Cherenkov resonance occurs when $V_b{\rm cos}\theta = \omega/k$, where $V_b$ is the laser group velocity or the beam velocity and $\theta$ is the angle between ${\textbf{\textit{V}}}_b$ and $\textbf{\textit{k}}$ \cite{Hutchinson1980,Yoshii1997}. Even for low-energy relativistic electrons, this condition can be easily satisfied. 
(ii) In the limit of $\omega_{p1}\gg\omega\gg\omega_{p2}, \Omega_c$,  
 $c^2k_x^2/\omega^2 \approx 1 - \omega_{p2}^2/\omega^2 + \omega^2/\omega_{p1}^2$.   
This branch intersects with the Cherenkov condition for forward-moving high-energy electrons around $\omega \simeq \omega_{H} = \sqrt{\omega_{p1}\omega_{p2}}$.

\begin{figure*}
	\includegraphics[width=18cm]{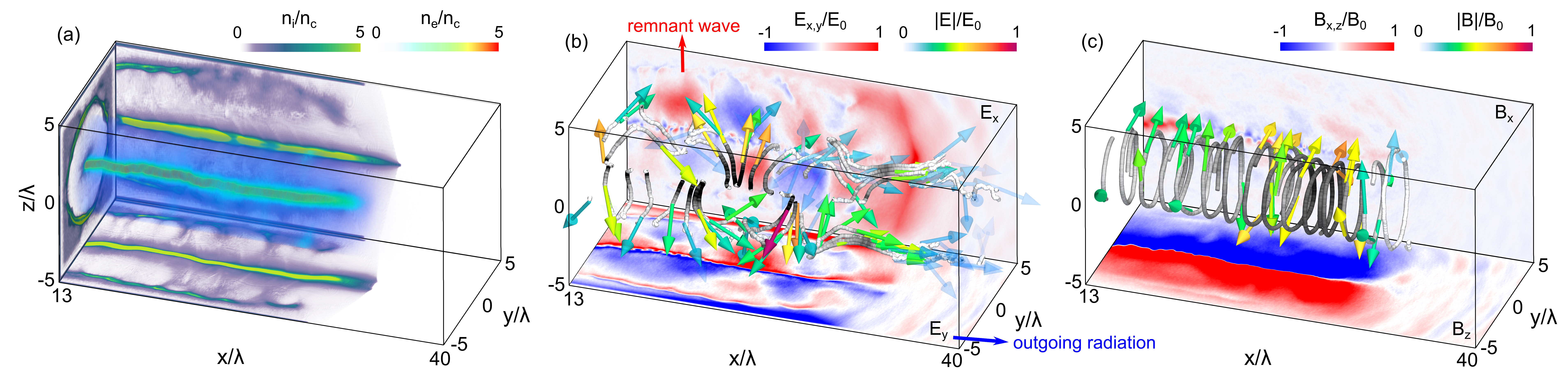}
	\caption{3D PIC simulation results. (a) Electron (volumetric) and ion (projections) density at $t=63T_0$. (b), (c) illustrate the electric and magnetic field components at $t=63T_0$, respectively, where the solid tubes represent the  field lines with their strength color coded (white to black) and directions indicated by the arrows. In (b), the back and bottom faces show the 2D cuts of the $E_x$ and $E_r$ (i.e. $E_y$) at $z=0$, respectively, while in (c), they correspondingly represent the 2D cuts of the $B_x$ and $B_\phi$ (i.e. $B_z$) at $z=0$. Here, $E_0,B_0=m_e\omega_0c/e$.
		\label{fig:fig2}
	}
\end{figure*}

To gain more insights into the dispersion relation, we numerically solve  Eq. (\ref{dispersion}). 
In Fig. \ref{fig:fig1}, we only show two key branches illustrated by the red lines. 
The low-frequency branch named C-mode is a new mode that is induced by the electron cyclotron motion. When $\omega\rightarrow\Omega_c$, $ck_x/\omega\rightarrow\infty$. For unmagnetized plasmas ($\Omega_c\rightarrow0$), this branch disappears and the dispersion relation turns to that of unmagnetized SPWs \cite{Shen2021,Macchi2018,Decyk1977}.  
When $\Omega_c>0$ and $\omega\rightarrow0$, $v_{ph}\rightarrow(\Omega_c/\omega_{u})c$ (black line), same as our above estimation. Furthermore, when $\omega\gg\omega_{p2},\,\Omega_c$, one can clearly see that the line of $v_{ph}=c$ intersects with the higher-frequency branch (H-mode) around $\omega = \omega_{H}$ (black dotted line), where the Cherenkov resonance can also occur.

In addition to the Cherenkov resonance, Doppler resonances have also been discussed for runaway electrons in tokamaks, where electrons mainly move along the magnetic field direction while performing cyclotron motion perpendicularly \cite{Martin2002,Kartashova2009}. The resonance occurs when the cyclotron frequency (or $n\Omega_c$) coincides with the wave frequency witnessed by the electron, i.e. $n\Omega_c + k_xv_x -\omega = 0$ in which $n$ is an integer. For $n>0$ ($n<0$), one has the normal (anomalous) Doppler resonance \cite{Martin2002}.  However, in our case, electrons move perpendicularly ($x$-direction) to the magnetic field direction while undergoing betatron oscillation in $y$-direction. This induces similar Doppler resonances and the condition can be given as
\begin{eqnarray}
	n\omega_\beta + k_xv_x -\omega = 0\label{Doppler},
\end{eqnarray}
where $\omega_\beta$ is betatron frequency. This condition is equivalent to the betatron resonance condition in direct laser acceleration, i.e. $n\omega_\beta = (1 - v_x/v_{ph})\omega$ \cite{Pukhov1999}. We show such Doppler resonances as the gray dashed lines in Fig. \ref{fig:fig1}, which may intersect with both the C- and H-modes. Note that here electrons can also undergo Doppler resonances with the excited mTPWs, not just the laser pulse.

To verify the above theory, we consider a scheme, that has been demonstrated in experiments and simulations to have advantages in the production of high-energy, high-flux particle beams \cite{Rosmej2020,Guenther2022,Ji2018,Gong2021}, where an intense linearly-polarized laser pulse interacts with a long NCD plasma. We perform 3D PIC simulations with the code \textsc{vlpl} \cite{Pukhov2016}, in which a
numerical-Cherenkov-free RIP Maxwell solver is used \cite{Pukhov2020}. A $y$-polarized laser pulse with intensity of $1.1\times10^{21}\,{\rm W/cm^2}$ and wavelength of $\lambda=1\mu$m impinges on a fully preionized uniform plasma with electron density $n_e = 1n_c$ and length $29\lambda$. Here, $n_c=m_e\omega_0^2/4\pi e^2$ is the critical density and $\omega_0$ is the laser frequency. The laser pulse has a Gaussian profile in both space and time with a focal spot diameter of $d_L=2.35\lambda$ and a pulse duration of $\tau_L=31.4$fs, respectively. The simulation box is $40\lambda\times10\lambda\times10\lambda$ in $x\times y\times z$-directions, containing $4000\times200\times200$ cells, respectively. The first $1\lambda$ and last $10\lambda$ space in $x$-direction are vacuum. We use four macroparticles per cell, both for electrons and carbon ions. Absorbing boundary conditions are used for particles and fields in each direction. We mention that we also observe similar phenomena in 2D simulations by using a $p$-polarized laser pulse.  

When the laser pulse channels into the NCD plasma, it cannot completely evacuate electrons from the channel and part of electrons may experience betatron resonance inside the channel, forming high-energy, high-density electron bunches \cite{Pukhov1999,Shen2021}.  
In our simulation, the density of accelerated electron bunches ($n_e>1.5n_c$) is higher than the initial plasma density, and therefore ions are accelerated radially inwards \cite{Popov2010}. This leads to the formation of a central filament with even higher density at later time, as shown in Fig. \ref{fig:fig2}(a), where the density of the central filament is larger than $10n_c$, much higher than the density of the surrounding plasma (about $0.3n_c$). Furthermore, during the LPI, the amplitude of the self-generated azimuthal field $B_\phi$  reaches about $0.5$ GG while the mean energies of the bulk and high-energy electrons are 5.6 MeV and 23 MeV, respectively. Thus, the corresponding cyclotron frequencies  $\Omega_c = eB_\phi/\gamma m_ec$  of these two groups of electrons are 0.4$\omega_0$ and 0.1$\omega_0$, and the gyroradii $r_c \simeq \gamma m_ec/eB_\phi$ are $0.5\lambda$ and $2\lambda$, respectively, 
much smaller than the time- and space-scale of interest here. This indicates that the plasma is strongly magnetized. Therefore, at the interface of the low-density channel plasma and high-density central plasma, mTPWs can be excited.

\begin{figure}
	\includegraphics[width=8.cm]{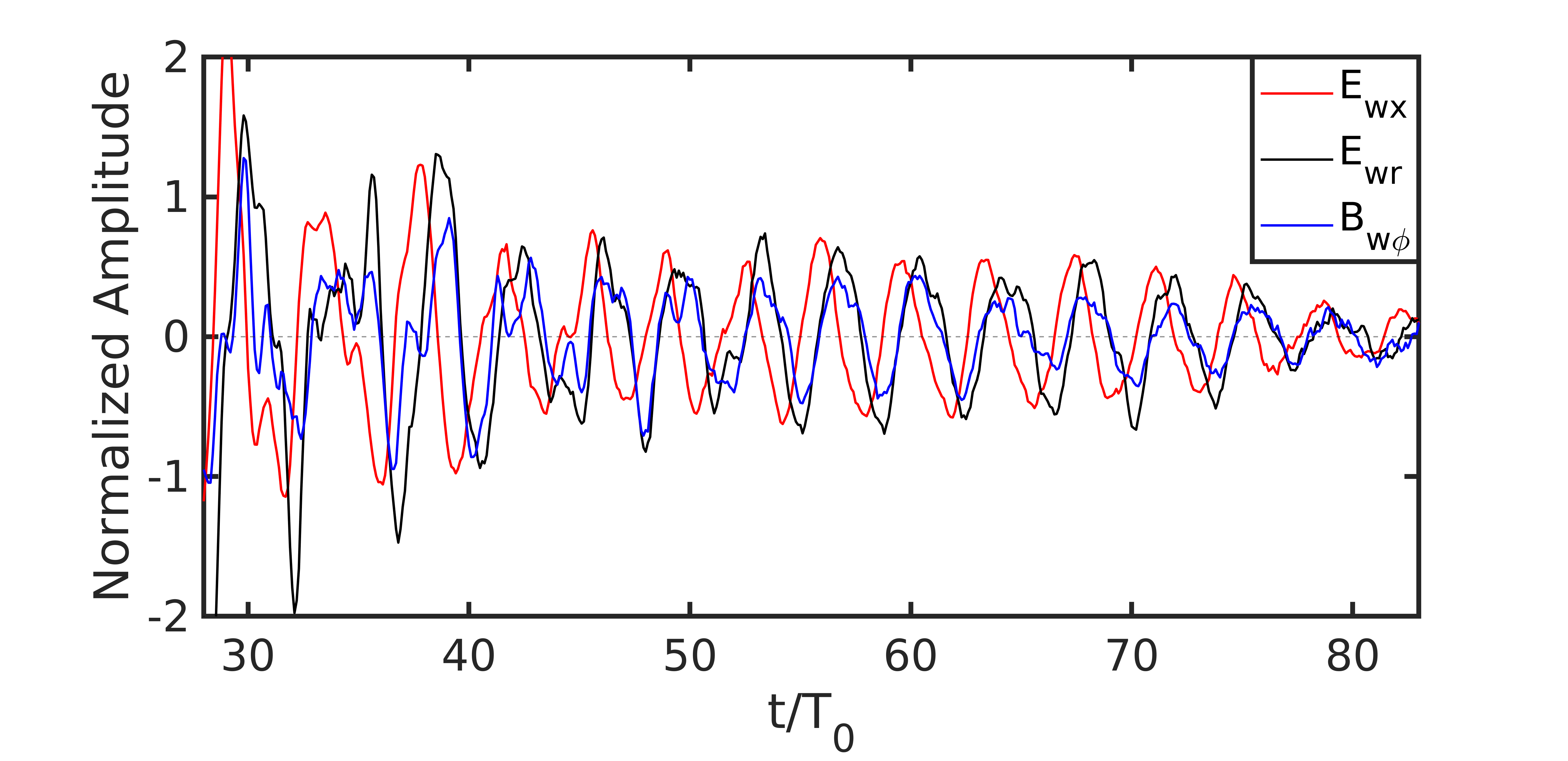}
	\caption{Temporal evolution of the field components of the mTPW registered at the point of ($16\lambda,1.05\lambda,0\lambda$). 
		\label{fig:fig3}
	}
\end{figure}

In Figs. \ref{fig:fig2}(b) and \ref{fig:fig2}(c) we present the field components of the magnetized wave at $t=63T_0$ when the laser pulse has already left the box, in which the solid tubes depicted the corresponding electric [\ref{fig:fig2}(b)] and magnetic [\ref{fig:fig2}(c)] field vectors with their strength color coded and directions indicated by the arrows. Here, $t=0$ represents the time the laser peak enters the plasma and $T_0 = 2\pi/\omega_0$. 
From Fig. \ref{fig:fig2}(b), we know that the electric field has both strong longitudinal $E_x$ and radial $E_r$ components. Their cuts at $z=0$ plane are correspondingly shown at the back and bottom faces, where one can clearly see the wave structure existing in both the $E_x$ and $E_y$ (i.e. $E_r$ at $z=0$) fields and their amplitudes are relativistically strong ($\sim 3\times10^{12}\,{\rm V/m}$). The wavelength of the remnant mode  inside the plasma along $x$-direction is about $8\lambda$. 
In Fig. \ref{fig:fig2}(c), it is evident that the magnetic field has a strong azimuthal $B_\phi$. However, because of the strong quasistatic field, the wave structure inside the plasma is not clear, see the longitudinal cut of $B_\phi$ field (i.e. $B_z$) shown at the bottom face.

To exclude the contribution of quasistatic fields and clearly show the field components $F_w$ of the excited mTPWs, we calculate $F_w = F - F_{s}$ where $F$ is the field recorded from simulations and $F_{s}$ represents the quasistatic field obtained by averaging $F$ over 8$T_0$. 
In Fig. \ref{fig:fig3}, we present the temporal evolution of the mTPW fields $E_{wx}$ (red), $E_{wr}$ (black) and $B_{w\phi}$ (blue), which are recorded at the point of ($16\lambda,1.05\lambda,0\lambda$) where we observe the strongest fields. The laser pulse passes through this point at $31T_0$. Soon thereafter, one can clearly see the fields are oscillating with a period about $4T_0$. The amplitudes of the $E_{wx}$, $E_{wr}$ and $B_{w\phi}$ fields are comparable and relativistically strong. They decrease very slowly, especially from $40T_0$ to $70T_0$.  
After that, due to the expansion of the ions, they start to decrease faster.

\begin{figure}
	\includegraphics[width=8.6cm]{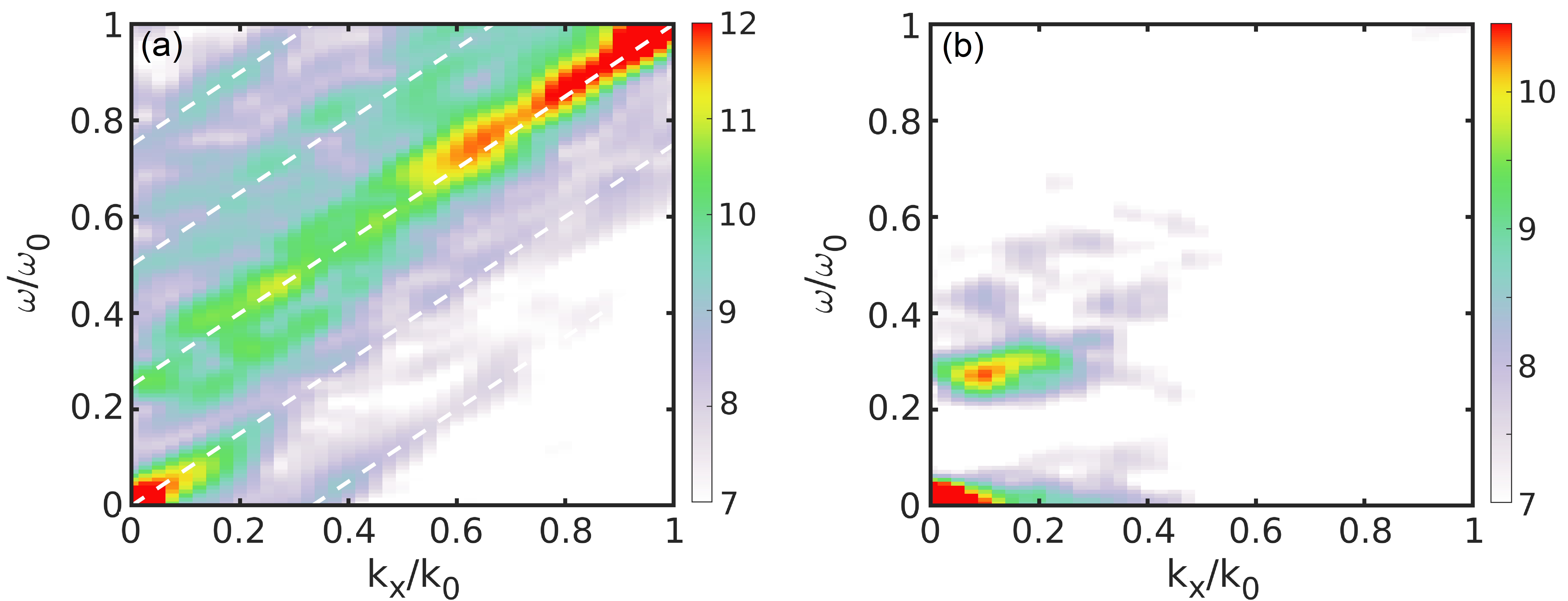}
	\caption{FFT of the $B_\phi$ field  at $t=23T_0$ (a) and $63T_0$ (b). Here, the field is registered at the line of ($y=1.05\lambda,z=0\lambda$) and a moving Gaussian window is applied to filter out the field. The white dashed lines in (a) represent the Doppler resonance  
	with $\omega_\beta = 0.25\omega_0$. A logarithmic color scale is used to show the modes clearly. 
		\label{fig:fig4}
	}
\end{figure}

To analyze the mode composition of the self-generated fields,  
we first record the temporal evolution of the magnetic fields at a line of ($y=1.05\lambda,z=0\lambda$), and then perform Fast Fourier Transform (FFT). The results are visualized in Fig. \ref{fig:fig4}. At $t=23T_0$, as shown in Fig. \ref{fig:fig4}(a), we know (i) There are two strong branches located at  domains of $(k_x\rightarrow0,\omega\rightarrow0)$ and $(k_x\rightarrow1,\omega\rightarrow1)$, corresponding to the Cherenkov resonance. This is in good agreement with our simple estimations from Eq. (\ref{dispersion}) and also the numerical solutions shown in Fig. \ref{fig:fig1}. (ii) There are several frequency-shifted branches almost parallel to each other, as marked by the white dashed lines in Fig. \ref{fig:fig4}(a), which are determined by the condition of Eq. (\ref{Doppler}) with $\omega_\beta = 0.25\omega_0$. Therefore, we deduce that they come from the normal and anomalous Doppler resonances.

\begin{figure}
	\includegraphics[width=8.cm]{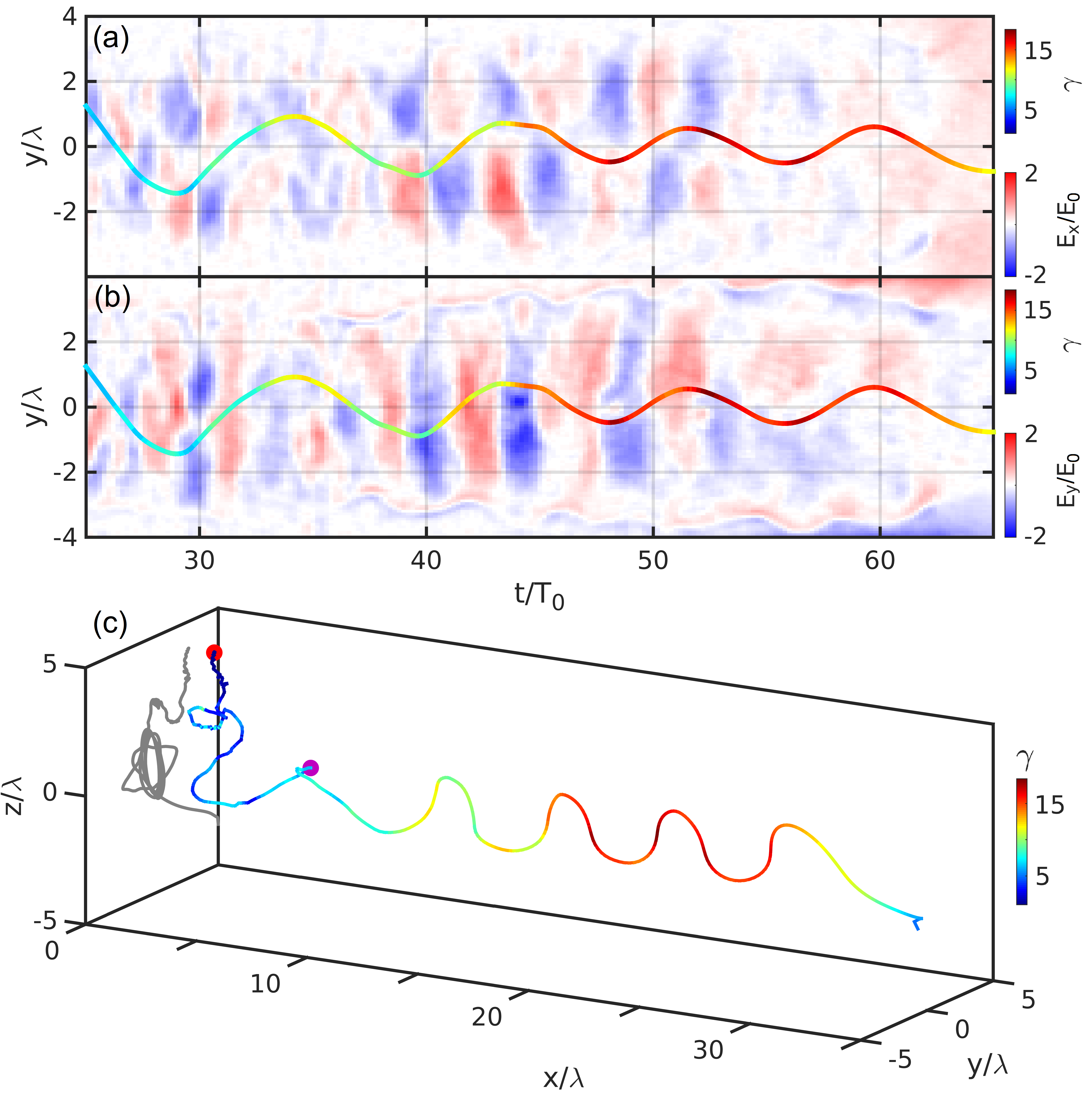}
	\caption{Electron acceleration in self-generated topological waves. (a), (b) show the  $E_x$ and  $E_y$ fields acting on the selected electron, respectively. (c) The 3D trajectory of the electron shows a quasihelical motion. The gray line in ($y,z$) plane represents its projection. The red dot marks its initial position and the purple marks its position at $t=25T_0$ corresponding to the start time shown in (a) and (b).
		\label{fig:fig5}
	}
\end{figure}

Most energetic electrons (up to 200 MeV) leave the plasma with the laser pulse at about $46T_0$. Subsequently, those relatively high frequency modes with relativistic amplitudes and wavelength about $3\mu$m also propagate through the plasma [see projections at the bottom face of Figs. \ref{fig:fig2}(b) and \ref{fig:fig2}(c)]. At $t=63T_0$, only a quasimonochromatic mode is left inside the plasma, as shown in Fig. \ref{fig:fig4}(b), since its group velocity is quite small (see Fig. \ref{fig:fig1}). Its frequency is close to the local plasma frequency of $0.27\omega_0$ and longitudinal wavenumber is about $0.1k_0$. This means that it is a fast mode with $v_{ph}\simeq2.7c$. From Figs. \ref{fig:fig2}(c) and \ref{fig:fig3}, we know that it is an electromagnetic mode that decays slowly. 

Depending on the sign of the change in momentum, under the Doppler resonance, electrons may feed energy into the wave (as we discussed before) or absorb energy from it. In the following, we show the process in which electrons are accelerated by the wave. The results are shown in Fig. \ref{fig:fig5}. Figures \ref{fig:fig5}(a) and \ref{fig:fig5}(b) show the $E_x$ and $E_y$ fields acting on a selected electron, respectively, where one can see that the electron undergoes transverse oscillations while moving forward. It gains energies from both the $E_x$ and $E_y$ fields. Its trajectory is depicted in Fig. \ref{fig:fig5}(c). An interesting phenomenon is that in 3D, its trajectory is quasihelical, instead of mainly oscillating in ($x,y$) plane. This might come from small $E_\phi$ and $B_r$ fields observed in our simulations since we used a $y$-polarized laser pulse that decides the interaction is not a perfect cylindrical symmetry.  
Such quasihelical motion will further induce a quasistatic $B_x$ field [back face of Fig. \ref{fig:fig2}(c)] that can in turn help trapping electrons and enhancing their radiation \cite{Liu2013}.

In conclusion, we analytically and numerically demonstrate the possibility of exciting magnetized topological surface waves in intense LPIs, specifically via a scheme where an intense femtosecond laser pulse interacts with a NCD plasma. We identify that such magnetized waves stem from the Cherenkov and Doppler resonances. The high-frequency part with relativistic amplitude and mid-infrared wavelength is transmitted into vacuum, while a long-living relativistic remnant mode  
further accelerates electrons inside plasma.   
Our findings pave the way for understanding the generation of magnetized topological waves in LPIs and their feedback effect on electron acceleration and radiation, and 
provide a potential approach to producing relativistic long-wavelength radiations, such as infrared waves \cite{Liu2010,Zhu2020}.

\section*{Acknowledgements}
This work is supported by the DFG (project PU 213/9). 
The authors gratefully acknowledge the Gauss Centre for Supercomputing e.V. (www.gauss-centre.eu) for funding this project by providing computing time through the John von Neumann Institute for Computing (NIC) on the GCS Supercomputer JUWELS at Jülich Supercomputing Centre (JSC). 
X.F.S. gratefully acknowledges support by the Alexander von Humboldt Foundation, as well as acknowledges helpful discussions with Prof. G. Shvets and Dr. R. S. Rajawat at Cornell University, Dr. G. Lehmann at HHU and Dr. K. Jiang at SZTU.


\end{document}